\documentclass[prd,showpacs,amsmath,amssymb]{revtex4}

\usepackage{graphicx}
\usepackage{dcolumn}
\usepackage{bm}
\usepackage{psfrag}
\usepackage{subfigure}

\newcommand{\be}{\begin{eqnarray}}
\newcommand{\ee}{\end{eqnarray}}
\newcommand{\bn}{\begin{eqnarray*}}
\newcommand{\en}{\end{eqnarray*}}

\newcommand{\nn}{\nonumber \\}
\newcommand{\nl}{\\}

\renewcommand{\vec}[1]{\mbox{\boldmath$#1$}}
\renewcommand{\d}{\mbox{\rm d}}

\newcommand{\rbar}{\ensuremath{\bar{r}}}
\renewcommand{\th}{\ensuremath{\theta}}
\newcommand{\ph}{\ensuremath{\varphi}}
\newcommand{\al}{\ensuremath{\alpha}}
\newcommand{\bt}{\ensuremath{\beta}}
\newcommand{\sg}{\ensuremath{\sigma}}
\newcommand{\gm}{\ensuremath{\gamma}}
\newcommand{\gfive}{\ensuremath{\gm^5}}

\newcommand{\pvec}{\ensuremath{\vec{p}}}
\newcommand{\Rvec}{\ensuremath{\vec{R}}}

\newcommand{\alvec}{\ensuremath{\vec{\al}}}
\newcommand{\sgvec}{\ensuremath{\vec{\sg}}}

\newcommand{\hb}{\ensuremath{\hbar}}
\newcommand{\ihb}{\ensuremath{i \hbar}}
\newcommand{\pt}[1]{\ensuremath{{\partial \over \partial #1}}}
\newcommand{\ptt}[1]{\ensuremath{{\partial^2 \over \partial {#1}^2}}}

\newcommand{\qp}{\ensuremath{{q \over p} \,}}
\newcommand{\ct}{\ensuremath{\rm CT}}
\newcommand{\nabvec}{\ensuremath{\vec{\nabla}}}
\newcommand{\PhiG}{\ensuremath{\Phi_{\rm G}}}

\newcommand{\lt}{\ensuremath{\left}}
\newcommand{\rt}{\ensuremath{\right}}
\newcommand{\dotprod}[2]{\ensuremath{\lt(\vec{#1} \cdot \vec{#2}\rt)}}
\newcommand{\cross}[2]{\ensuremath{\lt(\vec{#1} \times \vec{#2}\rt)}}

\renewcommand{\d}{\mbox{\rm d}}

\begin{document}

\pagenumbering{arabic}

\title{Neutrino Helicity and Chirality Transitions in Schwarzschild Space-Time}%

\author{Dinesh Singh}
\email{singhd@uregina.ca}
\affiliation{%
Department of Physics, University of Regina \\
Regina, Saskatchewan, S4S 0A2, Canada
}%
\date{\today}

\begin{abstract}

We study the helicity and chirality transitions of a high-energy neutrino propagating
in a Schwarzschild space-time background.
Using both traditional Schwarzschild and isotropic spherical co-ordinates, we
derive an ultrarelativistic approximation of the Dirac Hamiltonian to first-order
in the neutrino's rest mass, via a generalization of the Cini-Touschek transformation
that incorporates non-inertial frame effects due to the noncommutative nature of the
momentum states in curvilinear co-ordinates.
Under general conditions, we show that neutrino's helicity is not a constant of the motion
in the massless limit due to space-time curvature, while the chirality transition rate
still retains an overall dependence on mass.
We show that the chirality transition rate generally depends on the zeroth-order
component of the neutrino's helicity transition rate under the Cini-Touschek transformation.
It is also shown that the chiral current for high-energy neutrinos is altered by corrections
due to curvature and frame-dependent effects, but should have no significant bearing on
the chiral anomaly in curved space-time.
We determine the upper bound for helicity and chirality transitions near the
event horizon of a black hole.
The special case of a weak-field approximation is also considered, which includes the
gravitational analogue of Berry's phase first proposed by Cai and Papini.
Finally, we propose a method for estimating the absolute neutrino mass and
the number of right-handed chiral states from the expectation values
of the helicity and chirality transition rates in the weak-field limit.

\end{abstract}

\pacs{13.15.+g, 14.60.Lm, 04.20.-q, 04.70.Bw}

\maketitle

\setcounter{section}{0}
\setcounter{equation}{0}


\section{Introduction}

Recent experiments on solar neutrinos \cite{Kam,SNO} have provided strong evidence
that flavour oscillations occur while in transit to detectors on Earth, and suggest
that neutrinos have small but well-defined rest masses.
This suggestion implies that neutrinos can potentially change their helicity and chirality due
to interactions with external fields.
It is therefore possible to suppose that left-handed neutrinos can be converted to the
right-handed form, rendering them sterile to the weak interaction,
and therefore blind to all known forces except gravitation. 

Although a successful quantum theory of gravity currently does not exist,
for energy regimes and space-time length scales where quantum mechanics and
general relativity can safely appear together, the study of neutrinos in classical
gravitational backgrounds has the potential to yield, in principle, observationally
verifiable predictions.
For example, their helicity and chirality properties can act as probes of space-time
curvature effects appearing as quantum phase shifts to identify alternate theories
of gravity \cite{Adak}, while other neutrino properties may detect evidence of quantum
violations of the equivalence principle \cite{qviolation}, such as from interactions
with matter via the Mikheyev-Smirnov-Wolfenstein (MSW) mechanism \cite{Minakata}.
In addition, different aspects of neutrino dynamics may reveal valuable information
about the internal structure of neutron stars and supernovae \cite{Choudhury},
and the dark matter content of the known Universe \cite{Zuber}.
Because neutrinos are produced in abundance by stars and more exotic astrophysical
objects like neutron stars--whose surface gravity is especially strong--it is
important to have a better understanding of their spin and handedness properties due
to gravitational fields.

The purpose of this paper is to examine the helicity and chirality dynamics of a
high-energy massive neutrino while propagating in curved space-time.
A previous study \cite{Casini} of chiral transitions considered the special
case of weak gravitational fields.
It concluded that the transition amplitude is very small in a generally weak background,
but suggested that it may become large if the background space-time is strongly curved.
In this paper, we consider the same problem in the presence of strong gravitational fields, but
restrict attention to dynamics in a Schwarzschild space-time background.
An important feature of this approach is to use an ultrarelativistic form of the
Dirac Hamiltonian derived by the Cini-Touschek unitary transformation \cite{Cini},
which reasonably assumes that the neutrino rest mass is very small compared to its total energy.
When applied to a free particle in flat space-time, the corresponding Dirac Hamiltonian is
block diagonal in the chiral representation, where the mass-dependence is contained in the
energy eigenvalues, described in special relativistic form.
When applied to more general space-times, the mass terms in the Hamiltonian can be treated as
small first-order perturbations of the massless Dirac Hamiltonian, where we can effectively
control the degree of mass-dependence as required for a given problem.

We begin in \S 2 with a brief review of the formalism behind the Dirac equation in curved
space-time that leads to the corresponding Dirac Hamiltonian in general form,
followed by explicit expressions for Schwarzschild space-time as expressed in both regular
and isotropic spherical co-ordinates.
In \S 3, we use the derived Hamiltonians to determine the helicity and chirality transition rates for an
arbitrary fermion, to be later compared with equivalent expressions for the high-energy approximation
upon developing in \S 4 the formalism of the Cini-Touschek transformation in curvilinear
co-ordinates.
The helicity and chirality transitions for a high-energy neutrino are then determined explicitly in \S 5.
A detailed analysis of the results follows in \S 6, where special consideration is given to the case
of a weak-field approximation, which includes the gravitational analogue of Berry's phase first introduced
by Cai and Papini \cite{Cai}.
Furthermore, we examine the potential role played by the Cini-Touschek approximation on chiral symmetry
to determine its projected effect on the chiral anomaly in curved space-time.
We follow with a conclusion in \S 7, where possible future developments of these results are briefly explored.

For this paper, geometric units of $G = c = 1$ are assumed throughout, where we adopt
$-2$ signature for the metric.  General space-time indices are denoted by Greek characters and range from
0 to 3, while purely spatial indices are represented by Latin characters and range from 1 to 3.
However, when specifying a particular basis frame, the indices are labelled according to specific directions in space-time
with respect to co-ordinates which define the chosen frame.

\section{Dirac Hamiltonian in Curved Space-Time}
\setcounter{subsection}{0}

There is a considerable history on the study of spin-1/2 fermions in curved space-time \cite{Schmutzer}, where
both gravitational interactions and inertial forces due to accelerated motion can be introduced in a self-consistent
way \cite{Hehl1}.
The uniquely quantum mechanical nature of intrinsic spin angular momentum makes it a valuable probe of
quantum system properties in the presence of external classical fields, where predicted physical
effects occur for non-inertial fermion motion in Minkowski space-time \cite{Hehl2} that can be related to
gravitational effects via the principle of equivalence.
Already, a number of well-known experiments \cite{COW} suggest this correspondence can be effectively applied
here, although it must be stressed that critical issues of locality for locally accelerated systems \cite{Mashhoon}
may need to be carefully addressed when addressing quantum mechanical issues in curved space-time.
Nonetheless, there is no compelling reason to not follow this line of reasoning at this time.

\subsection{Covariant Dirac Equation}

The starting point here is the covariant Dirac equation,
\be
\lt[i \gamma^\mu (x) D_\mu - {m \over \hbar}\rt]\psi (x) & = & 0,
\label{covDirac=}
\ee
where $m$ is the neutrino rest mass and $D_\mu \equiv \hat{\nabla}_\mu + i \Gamma_\mu$
is the covariant derivative operator, with $\hat{\nabla}_\mu$ the usual
covariant derivative on index-labelled tensors and $\Gamma_\mu$ the spinor connection.
The gamma matrices $\lt\{\gamma^\mu(x) \rt\}$ satisfy
the conditions
$\lt\{\gamma^\mu (x), \gamma^\nu (x) \rt\} = 2 \, g^{\mu \nu}(x) $
and $D_\mu \, \gamma^\nu = 0 \,$.
Using a set of orthonormal tetrads $\lt\{\vec{e}_{\hat{\mu}} \rt\}$
and basis one-forms $\lt\{\vec{e}^{\hat{\mu}} \rt\}$
labelled by caratted indices to define a local Minkowski frame satisfying
$\lt\langle \vec{e}^{\hat{\mu}} , \vec{e}_{\hat{\nu}} \rt\rangle =
\delta^{\hat{\mu}}{}_{\hat{\nu}} \,$, the metric is described as
$\vec{g} = \eta_{\hat{\mu}\hat{\nu}} \, \vec{e}^{\hat{\mu}} \otimes
\vec{e}^{\hat{\nu}} \,$.
The general metric tensor and its Minkowski counterpart are related
by vierbein sets $\lt\{ e^{\hat{\alpha}}{}_\mu \rt\} \,$,
$\lt\{ e^\mu{}_{\hat{\alpha}} \rt\}$ satisfying
$\vec{e}^{\hat{\alpha}} = e^{\hat{\alpha}}{}_\beta \, \vec{e}^\beta$ and
$\vec{e}_{\hat{\alpha}} = e^\beta{}_{\hat{\alpha}} \, \vec{e}_\beta \,$, such that
\be
e^{\hat{\alpha}}{}_\mu \, e^\mu{}_{\hat{\beta}} & = &
\delta^{\hat{\alpha}}{}_{\hat{\beta}}\, , \quad
e^{\mu}{}_{\hat{\alpha}} \, e^{\hat{\alpha}}{}_{\nu} \ = \
\delta^\mu{}_\nu \, ,
\label{vierbein=}
\ee
and
\be
g_{\mu \nu} & = & \eta_{\hat{\alpha}\hat{\beta}} \,
e^{\hat{\alpha}}{}_\mu \, e^{\hat{\beta}}{}_\nu \, .
\label{gtensor=}
\ee
The spinor connection is
\be
\Gamma_\mu & = & -{1 \over 4} \, \sigma^{\alpha \beta}(x) \,
\Gamma_{\alpha \beta \mu} \ = \
-{1 \over 4} \, \sigma^{\hat{\alpha} \hat{\beta}} \,
\Gamma_{\hat{\alpha} \hat{\beta} \hat{\mu}} \, e^{\hat{\mu}}{}_\mu \, ,
\label{Gammadef=}
\ee
where $\sigma^{\hat{\alpha} \hat{\beta}} =
{i \over 2} [\gamma^{\hat{\alpha}}, \gamma^{\hat{\beta}}]$ are the
Minkowski space-time spin matrices, and $\Gamma_{\hat{\alpha} \hat{\beta} \hat{\mu}}$
are Ricci rotation coefficients derived from the Cartan equation of
differential forms
\be
\d\vec{e}^{\hat{\mu}} + \Gamma^{\hat{\mu}}{}_{\hat{\beta} \hat{\alpha}} \,
\vec{e}^{\hat{\alpha}} \wedge \vec{e}^{\hat{\beta}} & = & 0 \, .
\label{Cartan=}
\ee

\subsection{Dirac Hamiltonian in Schwarzschild Space-Time}

By arranging (\ref{covDirac=}) into the Schr\"{o}dinger form, it follows that
the Dirac Hamiltonian in general space-time co-ordinates is
$i \hbar \, \pt{t} \, \psi(x) = H \, \psi(x)$, where
\be
H & = & \lt(g^{00}\rt)^{-1} \, e^0{}_{\hat{\mu}} \lt[
\gamma^{\hat{\mu}} \, m + e^j{}_{\hat{\nu}} \lt(\eta^{\hat{\mu} \hat{\nu}}
- i \, \sigma^{\hat{\mu} \hat{\nu}} \rt) \lt(-i \hbar \, \partial_j +
\hbar \, \Gamma_j \rt) \rt] + \hbar \, \Gamma_0 \, .
\label{Hgeneral=}
\ee
We can now consider a derivation of the Hamiltonian for Schwarzschild space-time.
For the metric in Schwarzschild co-ordinates, the orthonormal basis frame is
\be
\vec{e}^t & = &  F(r) \, \d t \, , \quad
\vec{e}^r \ = \ {1 \over F(r)} \, \d r \, , \quad
\vec{e}^\th \ = \ r \, \d \theta \, , \quad
\vec{e}^\ph \ = \ r \, \sin \theta \, \d \phi \, ,
\label{1-form-Schw=}
\ee
where $F(r) = \lt(1 - 2M/r\rt)^{1/2}$ is the lapse function.
The Dirac Hamiltonian (\ref{Hgeneral=}) is then
\be
H & = & F\lt[m \, \bt + \dotprod{\al}{p}\rt] + F \lt(F - 1\rt) \alvec^r \pvec^r
\nn
& &{}- \ihb \, {F \over r} \lt[F \lt(1 + r \, \pt{r} \ln F^{1/2} \rt) \alvec^r + {1 \over 2} \, \cot \th \, \alvec^\th \rt] \, ,
\label{H-Schw=}
\ee
where $\pvec = -\ihb \, \nabvec$ in spherical co-ordinates
is
\be
\pvec^r & = & - \ihb \, \pt{r} \, , \quad
\pvec^\th \ = \ - {\ihb \over r} \, \pt{\th} \, , \quad
\pvec^\ph \ = \ - {\ihb \over r \, \sin \th} \, \pt{\ph} \, ,
\label{momentum1=}
\ee
and the dot product operation denoted in (\ref{H-Schw=}) and elsewhere in this paper is
$\dotprod{A}{B} \equiv \vec{A}^{\hat{1}} \vec{B}^{\hat{1}} + \vec{A}^{\hat{2}} \vec{B}^{\hat{2}}
+ \vec{A}^{\hat{3}} \vec{B}^{\hat{3}} \,$.

For Schwarzschild space-time written in isotropic spherical co-ordinates \cite{Carmeli},
the one-form frame corresponding to the metric is
\be
\vec{e}^t & = & \lt({R_-(\rbar) \over R_+(\rbar)}\rt) \, \d t \, , \quad
\vec{e}^{\rbar} \ = \ \lt(R_+(\rbar)\rt)^2 \, \d \rbar \, , \quad
\vec{e}^\th \ = \ \lt(R_+(\rbar)\rt)^2 \, \rbar \, \d \th \, , \quad
\vec{e}^\ph \ = \ \lt(R_+(\rbar)\rt)^2 \, \rbar \, \sin \th \, \d \ph \, ,
\label{1-form-isotropic=}
\ee
where $R_\pm(\rbar) = 1 \pm M/2 \, \rbar$, and
\be
\rbar(r) & = & {r \over 2} \, \lt[\lt(F(r)\rt)^2 + F(r) + {M \over r} \rt] \, ,
\nn
r(\rbar) & = & \lt(R_+(\rbar)\rt)^2 \, \rbar \, .
\label{mapping=}
\ee
Then the Dirac Hamiltonian in isotropic co-ordinates is
\be
H & = & {R_- \over R_+^3} \lt[m \, R_+^2 \, \bt +  \dotprod{\al}{p}
- {\ihb \over \rbar} \lt[\lt(1 + \rbar \pt{\rbar} \ln \lt(R_- R_+^3 \rt)^{1/2}\rt) \alvec^{\rbar}
+ {1 \over 2} \cot \th \, \alvec^\th\rt] \rt] \, ,
\label{H-Schw-isotropic=}
\ee
where $\vec{p}$ is described by
\be
\pvec^{\rbar} & = &  - \ihb \, \pt{\rbar}
\ = \ - \ihb \, {2 \, r^2 \over M^2} \,
F(r) \lt[\lt(F(r)\rt)^2 + F(r) + {M \over r} \rt] \pt{r} \, ,
\nn
\pvec^\th & = &  -{\ihb \over \rbar} \, \pt{\th} \, , \quad
\pvec^\ph \ = \ -{\ihb \over \rbar \, \sin \th} \, \pt{\ph} \, .
\label{Pisotropic=}
\ee

\section{Helicity and Chirality Transitions in Schwarzschild Space-Time}

\subsection{Non-Inertial Dipole Operator in Curvilinear Co-ordinates}

Having now the Dirac Hamiltonians (\ref{H-Schw=}) and (\ref{H-Schw-isotropic=}) in Schwarzschild and
isotropic spherical co-ordinates, respectively, we can proceed to determine
the helicity and chirality transition rates for an arbitrary uncharged fermion of mass $m$.
However, before proceeding it is necessary to first recognize that the momentum operators
(\ref{momentum1=}) and (\ref{Pisotropic=}) have co-ordinate dependence, which render them {\em non-commutative}.
That is, for a momentum operator in general curvilinear co-ordinates
\be
\pvec^{\hat{\imath}} & = &  - \ihb \nabvec_{\hat{\imath}}
\ = \ - {\ihb \over \lambda^{\hat{\imath}}(u)} \, \pt{u^{\hat{\imath}}} \, ,
\label{momentum-curve=}
\ee
where $\lambda^{\hat{\imath}}(u)$ are the scale functions corresponding to the given co-ordinate system,
then it is generally true that $\lt(i/\hb\rt)\lt[\pvec^{\hat{\imath}}, \pvec^{\hat{\jmath}}\rt] \equiv
N^{\hat{\imath}\hat{\jmath}} \neq 0 \,$.
Given the three-dimensional Levi-Civita alternating symbol $\epsilon^{ijk}$ with indices raised and
lowered by the Minkowski metric, and where $\epsilon^{123} \equiv 1$, we can introduce \cite{Singh3} a new
Hermitian vector operator $\Rvec \,$, which we call a {\em non-inertial dipole operator}, for reasons that will be
clear later in this paper.
The components for $\Rvec$ in general curvilinear co-ordinates are then
\be
\Rvec^{\hat{k}} & = & {i \over 2\hbar} \, \epsilon_{ij}{}^k \, [\pvec^{\hat{\imath}}, \pvec^{\hat{\jmath}}]
\ = \ {1 \over 2} \, \epsilon_{ij}{}^k \, N^{\hat{\imath}{\hat{\jmath}}}
\nn
& = & \delta^k{}_1 \left[{1 \over \lambda^{\hat{3}}(u)}
\left({\partial \over \partial u^{\hat{3}}} \, \ln \lambda^{\hat{2}}(u) \right) \pvec^{\hat{2}} -
{1 \over \lambda^{\hat{2}}(u)}
\left({\partial \over \partial u^{\hat{2}}} \, \ln \lambda^{\hat{3}}(u) \right) \pvec^{\hat{3}}
\right]
\nn
&  &{} + \delta^k{}_2 \left[{1 \over \lambda^{\hat{1}}(u)}
\left({\partial \over \partial u^{\hat{1}}} \, \ln \lambda^{\hat{3}}(u) \right) \pvec^{\hat{3}} -
{1 \over \lambda^{\hat{3}}(u)}
\left({\partial \over \partial u^{\hat{3}}} \, \ln \lambda^{\hat{1}}(u) \right) \pvec^{\hat{1}}
\right]
\nn
&  &{} + \delta^k{}_3 \left[{1 \over \lambda^{\hat{2}}(u)}
\left({\partial \over \partial u^{\hat{2}}} \, \ln \lambda^{\hat{1}}(u) \right) \pvec^{\hat{1}} -
{1 \over \lambda^{\hat{1}}(u)}
\left({\partial \over \partial u^{\hat{1}}} \, \ln \lambda^{\hat{2}}(u) \right) \pvec^{\hat{2}}
\right] \, ,
\label{Rvec=}
\ee
where it is obvious that $\Rvec$ vanishes identically for a Cartesian co-ordinate system.
For the momentum operators (\ref{momentum1=}) and (\ref{Pisotropic=}), it follows from (\ref{Rvec=}) that
\begin{subequations}
\be
\Rvec^r & = & {i \over \hb}[\pvec^\th, \pvec^\ph] \ = \
- {\cot \th \over r} \, \pvec^\ph \, ,
\nl
\Rvec^\th & = & {i \over \hb}[\pvec^\ph, \pvec^r] \ = \
{1 \over r} \, \pvec^\ph \, ,
\nl
\Rvec^\ph & = & {i \over \hb}[\pvec^r, \pvec^\th] \ = \
- {1 \over r} \, \pvec^\th \, ,
\label{Rvec1=}
\ee
\end{subequations}
for Schwarzschild co-ordinates, and
\begin{subequations}
\be
\Rvec^{\rbar} & = & {i \over \hb}[\pvec^\th, \pvec^\ph] \ = \
- {\cot \th \over \rbar} \, \pvec^\ph \, ,
\nl
\Rvec^\th & = & {i \over \hb}[\pvec^\ph, \pvec^{\rbar}] \ = \
{1 \over \rbar} \, \pvec^\ph \, ,
\nl
\Rvec^\ph & = & {i \over \hb}[\pvec^{\rbar}, \pvec^\th] \ = \
- {1 \over \rbar} \, \pvec^\th \, ,
\ee
\label{Rvec2=}
\end{subequations}
for isotropic spherical co-ordinates.
It is clear from both representations that $\Rvec$ induces some form of apparent rotational effect with
an associated Lie algebra, due to the mixing of radial and angular momentum states.
As well, for a fermion propagating solely in the radial direction, $\Rvec$ vanishes.

\subsection{Helicity and Chirality Transition Rates}

Having now determined $\Rvec$ explicitly for the Schwarzschild and isotropic spherical co-ordinate systems employed here,
it is straightforward to compute the helicity transition rate of an arbitrary fermion, as defined by $h \equiv \dotprod{\sg}{p}$.
Adopting the notation
\be
\cross{A}{B}^{\hat{\imath}} & \equiv & \epsilon^i{}_{jk} \, \vec{A}^{\hat{\jmath}} \, \vec{B}^{\hat{k}} \, ,
\label{crossprod=}
\ee
where it is understood that (\ref{crossprod=}) only truly represents the ith component of a three-dimensional cross product
when applied to Cartesian co-ordinates, it follows from the Heisenberg equations of motion that for an observer very far from
the gravitational source,
\be
\dot{h} & = & F \lt\{- m \lt(\pt{r} \ln F \rt) \bt \, \sgvec^r + {2 \over \hb} \lt(F - 1\rt) \cross{\al}{p}^r \pvec^r
- F \lt(\pt{r} \ln F^2 \rt) \pvec^r + {\hb \over 2 \, r^2} \, \cot \th \lt(1 - r \pt{r} \ln F \rt) \alvec^\ph \rt.
\nn
& &{}
- {2 \, i \over r} \lt[\lt(F + \lt(F - 1\rt) r \pt{r} \ln F^{1/2}\rt)\cross{\al}{p}^r + {1 \over 2} \, \cot \th \cross{\al}{p}^\th \rt]
+ i \lt(F - 1\rt) \lt[\dotprod{\al}{R} - \alvec^r \Rvec^r \rt]
\nn
& &{} + \lt. \ihb \, \gfive \lt[F \lt[\lt(\pt{r} \ln F\rt)^2 + \ptt{r} \ln F^{1/2}\rt]
- {F \over r^2} \lt(1 - r \pt{r} \ln F^2\rt) - {1 \over 2 \, r^2} \, {1 \over \sin^2 \th} \rt] \rt\}
\label{hdot-Sch-Ortho}
\ee
for Schwarzschild co-ordinates, and
\be
\dot{h} & = & {R_- \over R_+^3} \lt\{
- m \, R_+^2 \, \pt{\rbar} \ln \lt({R_- \over R_+}\rt) \bt \, \sgvec^{\rbar}
- \pt{\rbar} \ln \lt({R_- \over R_+^3}\rt) \pvec^{\rbar}
+ {\hb \over 2 \, \rbar^2} \, \cot \th \lt[1 - \rbar \pt{\rbar} \ln \lt({R_- \over R_+^3}\rt) \rt] \alvec^\ph
 \rt.
\nn
& &{} - {2 \, i \over \rbar} \lt[
\lt(1 + \rbar \pt{\rbar} \ln R_+^3\rt)\cross{\al}{p}^{\rbar} + {1 \over 2} \, \cot \th \cross{\al}{p}^\th \rt]
\nn
& &{} + \lt. \ihb \, \gfive
\lt[\pt{\rbar} \ln \lt({R_- \over R_+^3}\rt) \pt{\rbar} \ln \lt(R_- R_+^3\rt)^{1/2}
+ \ptt{\rbar} \ln \lt(R_- R_+^3\rt)^{1/2} 
- {1 \over \rbar^2}
\lt[1 - \rbar \pt{\rbar} \ln \lt(R_- R_+^3\rt)\rt] - {1 \over 2 \, \rbar^2} \, {1 \over \sin^2 \th}\rt] \rt\}
\nn
\label{hdot-Sch-Iso}
\ee
for isotropic spherical co-ordinates.

Focussing on the Hermitian components of (\ref{hdot-Sch-Ortho}) and (\ref{hdot-Sch-Iso}), we notice a parallel
set of contributions to the helicity transition rate in both Schwarzschild and isotropic spherical co-ordinates.
There is a noticable exception due to presence of the second term in (\ref{hdot-Sch-Ortho}) not found in
(\ref{hdot-Sch-Iso}).
However, this term is non-zero only for propagation that is not purely radial with respect to the spatial
origin of the space-time background.
When this is taken into account, the correspondence between the two expressions is more complete.
Apart from the co-ordinate singularity for $\th = 0$ and $\th = \pi$, the helicity transition rate is finite
for fermion propagation in vacuum away from a regular non-rotating star.
For the special case of a neutron star or black hole, it turns out that while the transition rate becomes
singular for a fermion near the event horizon in Schwarzschild co-ordinates due to a gravitational blueshift
effect, we obtain a finite expression in isotropic co-ordinates, where
\be
\lt. \dot{h} \rt|_{\rbar \approx M/2} & \approx & -{m \over M} \, \bt \, \sgvec^{\rbar} - {1 \over 4M} \, \pvec^{\rbar}
- {\hb \over 4M^2} \, \cot \th \, \alvec^\ph \, .
\label{hdot-Sch-Iso-EH}
\ee
Given the $M^{-1}$ dependence on (\ref{hdot-Sch-Iso-EH}), a helicity transition for an arbitrary fermion is
essentially nonexistent for astrophysically large black holes and neutron stars, but may be significant for
microscopic black holes.

On a more theoretical level, (\ref{hdot-Sch-Ortho}) and (\ref{hdot-Sch-Iso}) make an important suggestion that,
in the limit as $m \rightarrow 0$, the fermion's helicity transition rate is {\em not} a constant of the
motion, due to its interaction with the space-time metric during propagation.
This claim has also been put forward \cite{Singh3} in the context of accelerated fermion beams in circular
storage rings, where the external magnetic field required to bend the beam generates the violation of
helicity conservation.
Therefore, it should not be too surprising to obtain a similar result due to a non-trivial classical
space-time background.
In sharp contrast, the chirality transition rate is a constant of the motion in the massless limit, since
for the projection operator $P_\pm \equiv \lt(1 \pm \gfive\rt)/2$ where the upper sign refers to the right-handed
state,
%
%
\begin{subequations}
\be
\dot{P}_\pm & = & \pm {i \over \hb} \, m \, F \, \bt \, \gfive \, , \quad {\rm (Schwarzschild)}
\nl
& = & \pm {i \over \hb} \, m \, {R_- \over R_+} \, \bt \, \gfive \, , \quad {\rm (Isotropic)}
\ee
\end{subequations}
which has an overall mass dependence.
Furthermore, it is known that for \cite{Casini}
\be
\lt| \psi_{\rm C} \rt\rangle & = & C_{\rm R} \lt| \ph_{\rm R} \rt\rangle + C_{\rm L} \lt| \ph_{\rm L} \rt\rangle \, ,
\label{psi-chiral}
\ee
where
\be
\lt| \ph_{\rm R} \rt\rangle & = &
\lt(\begin{array}{c}
\ph_{\rm R} \\
0
\end{array}\rt) \, , \qquad
\lt| \ph_{\rm L} \rt\rangle \ = \
\lt(\begin{array}{c}
0 \\
\ph_{\rm L}
\end{array}
 \rt) \, ,
\label{chiral}
\ee
are the right- and left-handed chiral states
subject to the normalization condition $\lt|C_{\rm R}\rt|^2 + \lt|C_{\rm L}\rt|^2 = 1$,
it follows that $\lt\langle \bt \gfive \rt\rangle \equiv \lt\langle \psi_{\rm C} \rt| \bt \gfive \lt| \psi_{\rm C} \rt\rangle = 0$.
That is, the chirality remains a constant of the motion, even for massive fermions.
This discrepancy challenges the well-known identification \cite{Itzykson} between helicity and chirality
for massless fermions, and merits further investigation.


\section{Cini-Touschek Transformation and the High-Energy Dirac Hamiltonian in Schwarzschild Space-Time}

Although (\ref{hdot-Sch-Ortho}) and (\ref{hdot-Sch-Iso}) are obviously useful as exact expressions for the helicity transition
rate of an arbitrary uncharged fermion, it is also useful to reconsider the problem in terms of a high-energy
approximation of the Dirac Hamiltonian, which can be written as a power series expansion with respect to $q \equiv m/p \ll 1$.
By doing this, we can identify the leading-order terms that contribute to the dynamics
of particles which satisfy this condition.
High-energy massive neutrinos certainly qualify for this treatment.
In addition, by adopting the chiral representation \cite{Itzykson}
\be
\beta & = &
\lt( \begin{array}{cc}
0 & -1 \\
-1 & 0
\end{array} \rt) \, ,
\qquad
\vec{\alpha}^{\hat{\imath}} \ = \
\lt( \begin{array}{cc}
\vec{\sigma}^{\hat{\imath}} & 0 \\
0 & -\vec{\sigma}^{\hat{\imath}}
\end{array} \rt) \, ,
\qquad
\gamma^5 \ = \
\lt( \begin{array}{cc}
1 & 0 \\
0 & -1
\end{array} \rt),
\label{chiral=}
\ee
we seek to re-express the Hamiltonian such that the zeroth-order term is in
block diagonal form (while still retaining a particle mass dependence), where the
first-order term is a mass perturbation.
This allows for an effective decoupling of the four-spinor wavefunction into its
right- and left-handed states,
while the perturbation term is responsible for their mixing.
It is known that, by applying the Cini-Touschek (CT) transformation to the
free-particle Hamiltonian \cite{Cini}, the new Hamiltonian effectively removes the
$\beta$-matrix term and yields an energy eigenvalue $E \ = \ \pm \sqrt{p^2 + m^2}$.

In this Section, we re-derive the CT transformation for the free-particle Hamiltonian
in curvilinear co-ordinates, then use it to evaluate the high-energy form of the
Hamiltonian in Schwarzschild space-time and its weak-field limit.
Unlike the more well-known Foldy-Wouthuysen (FW) transformation \cite{Greiner}, used extensively
to model the non-relativistic motion of a spin-1/2 particle in external fields, the CT
transformation has not been as widely used for similar purposes.
A preliminary study has recently been considered \cite{Singh} in the context of non-inertial
motion through Minkowski space-time, where the CT transformation is performed
on a local inertial tangent space in a frame comoving with the particle, with the outcome
described by general co-ordinates after projecting back onto the space-time manifold.
The approach taken here instead follows the known procedure when applied to the FW transformation,
as adopted earlier \cite{Casini} in the study of neutrinos in weak gravitational fields.


\subsection{Free-Particle High-Energy Hamiltonian in Curvilinear Co-ordinates}

To derive the CT transformation for the free-particle Dirac Hamiltonian \cite{Singh3}, we begin with the
unitary operator $\exp\lt(i S_{\rm CT}\rt)$, where
\be
S_{\rm CT} & = & {i \, \omega(q) \over 2p} \, \beta \dotprod{\al}{p} \, ,
\label{Sct=}
\ee
and $\omega(q)$ is a constraint function used to absorb the $\beta$ matrix.
Then
\be
H_{\rm CT} & = & e^{i S_{\rm CT}} \lt[m \beta +
\lt(\vec{\alpha} \cdot \vec{p}\rt) \rt] e^{-i S_{\rm CT}}
\ = \
e^{2i S_{\rm CT}} \lt[m \beta + \lt(\vec{\alpha} \cdot \vec{p}\rt) \rt] \, .
\label{Hct1=}
\ee
By Taylor expansion,
\be
e^{2i S_{\rm CT}} & = & 1 - {\omega \over p} \, \beta \dotprod{\al}{p}
+ {1 \over 2!} \lt[{\omega \over p} \, \beta \dotprod{\al}{p}\rt]^2
- {1 \over 3!} \lt[{\omega \over p} \, \beta \dotprod{\al}{p}\rt]^3
+ \cdots \, ,
\label{exp1=}
\nl
\lt[{\omega \over p} \, \beta \dotprod{\al}{p}\rt]^2
& = & - {\omega^2 \over p^2} \dotprod{\al}{p}^2
\ = \ - {\omega^2 \over p^2} \lt[\dotprod{p}{p}
- {i \over 2} \,  \epsilon_{ijk} \, \sgvec^{\hat{\imath}} \, [\pvec^{\hat{\jmath}}, \pvec^{\hat{k}}]\rt]
\nn
& = & - {\omega^2 \over p^2} \lt[\dotprod{p}{p} + \hb \, \dotprod{\sg}{R}\rt]\, .
\label{exp2=}
\ee
With this expression, $\vec{\sigma} \cdot \vec{R}$ is analogous to a magnetic dipole term
due to the curl of the electromagnetic vector potential, hence the reference to $\Rvec$ as a
non-inertial dipole operator.

By substituting (\ref{exp2=}) into (\ref{exp1=}), we show that
\be
e^{2i S_{\rm CT}} & \approx & \cos\lt[\omega \lt(1 + {\hb \over 2p^2} \, \dotprod{\sg}{R} \rt) \rt]
- \sin\lt[\omega \lt(1 + {\hb \over 2p^2} \, \dotprod{\sg}{R} \rt) \rt]\lt(1 - {\hb \over 2p^2} \,
\dotprod{\sg}{R} \rt) \beta {\dotprod{\al}{p} \over p} \, .
\label{exp3=}
\ee
Applying (\ref{exp3=}) to (\ref{Hct1=}) and setting the coefficient of $\beta$ to zero,
it follows that
\be
\omega(q) & \approx & \lt(1 - {\hbar \over 2p^2} \, \dotprod{\sg}{R} \rt)
\tan^{-1} \lt[q \lt(1 - {\hbar \over 2p^2} \, \dotprod{\sg}{R} \rt) \rt] \, ,
\label{omega=}
\ee
with the final result that
\be
e^{i S_{\rm CT}} \lt[m \beta + \dotprod{\al}{p} \rt]
e^{-i S_{\rm CT}} & \approx & \lt[\sqrt{p^2 + m^2} - {q^3 \over \sqrt{1 + q^2}}
\, {\hbar \over 2m} \, \dotprod{\sg}{R} \rt] {\dotprod{\al}{p} \over p} \, .
\label{Hct2=}
\ee
For the small angle approximation of (\ref{omega=}),
\be
\omega(q) & \approx & q \lt(1 - {\hbar \over p^2} \, \dotprod{\sg}{R} \rt).
\label{omega1=}
\ee

In Minkowski space-time, the presence of the dipole term in (\ref{Hct2=}) suggests that a small
energy shift appears for a spin-1/2 particle in the ultrarelativistic approximation.
For high-energy neutrinos, this shift should be negligible compared to its relativistic energy
$\sqrt{p^2 + m^2}$, since the dipole term is order $q^3 \ll \ll 1$.
However, for neutrinos with lower energies and other spin-1/2 particles with larger rest
masses, it may be a detectable effect for large enough $q$.

\subsection{High-Energy Hamiltonian}

To derive the high-energy form of some operator using the
CT transformation, we use the series expansion
\cite{Greiner}
\be
X_{\rm CT} & = & e^{i S_{\rm CT}} \, X \, e^{-i S_{\rm CT}} \ = \
X + i [S_{\rm CT}, X] + {i^2 \over 2!} \lt[S_{\rm CT}, [S_{\rm CT}, X] \rt]
+ {i^3 \over 3!} \lt[S_{\rm CT}, \lt[S_{\rm CT}, [S_{\rm CT}, X] \rt]\rt] + \cdots,
\label{Xexpand1=}
\ee
where $S_{\rm CT}$ is of order $q$ and $X$ is an arbitrary operator.
Because the CT transformation operator is unitary, the eigenvalues associated
with the transformed operator remain unchanged \cite{Sakurai}.
For $q$ sufficiently small, as satisfied by the case for massive neutrinos, it is
sufficient to include only the first-order expansion term in (\ref{Xexpand1=}).
Therefore,
\be
e^{i S_{\rm CT}} \, X \, e^{-i S_{\rm CT}} & \approx &
X - {\omega(q) \over 2p} \lt[ \beta \dotprod{\al}{p}, X \rt].
\label{Xexpand2=}
\ee
We therefore insert $1 = e^{i S_{\rm CT}} \, e^{-i S_{\rm CT}}$
between each factor and operator that comprises the given Hamiltonian and evaluate to
first-order in $\omega(q)$ using (\ref{Xexpand2=}), keeping terms only up to first-order in $\hb$.

By applying (\ref{Hct2=}), (\ref{omega1=}), and (\ref{Xexpand2=}) for $X$ replaced by $H$, it follows that
\be
H_{\ct} & \approx & F \lt\{\dotprod{\al}{p} + \lt(F - 1\rt) \alvec^r \pvec^r
- {\ihb \over r} \lt[F \lt(1 + r \, \pt{r} \ln F^{1/2} \rt) \alvec^r + {1 \over 2} \, \cot \th \, \alvec^\th \rt] \rt.
\nn
& &{} + \qp \bt \lt[-\lt(F - 1\rt) \lt(1 - {\hb \over p^2} \dotprod{\sg}{R} \rt)\pvec^r \pvec^r
+ {\hb \over 2} \lt[\pt{r} \ln F \cross{\sg}{p}^r - \lt(F - 1\rt)\lt[\dotprod{\sg}{R} - \sgvec^r \Rvec^r \rt] \rt] \rt]
\nn
& &{} + \lt. {1 \over 2} \, q^2 \dotprod{\al}{p}
- {q^3 \over p} \, {\hb^2 \over 2 \,m} \lt[\gfive \dotprod{R}{p} + i \, \alvec \cdot \cross{R}{p} \rt] \rt\}
\label{HCT-Sch-Ortho}
\ee
for Schwarzschild co-ordinates, and
\be
H_{\ct} & \approx & {R_- \over R_+^3} \lt\{ \dotprod{\al}{p}
- {\ihb \over \rbar} \lt[\lt(1 + \rbar \pt{\rbar} \ln \lt(R_- R_+^3 \rt)^{1/2}\rt) \alvec^{\rbar}
+ {1 \over 2} \cot \th \, \alvec^\th\rt] \rt.
\nn
& &{} + \qp \bt
\lt[\lt(R_+^2 - 1\rt) \lt(1 - {\hb \over p^2} \dotprod{\sg}{R} \rt)\dotprod{p}{p}
+ \hb \lt[\pt{\rbar} \ln \lt(R_- \over R_+\rt)^{1/2} \cross{\sg}{p}^{\rbar} + \dotprod{\sg}{R}\rt] \rt]
\nn
& &{} + \lt. R_+^2 \lt[{1 \over 2} \, q^2 \, \dotprod{\al}{p}
- {q^3 \over p} \, {\hb^2 \over 2 \,m} \lt[\gfive \dotprod{R}{p} + i \, \alvec \cdot \cross{R}{p} \rt] \rt] \rt\}
\label{HCT-Sch-Iso}
\ee
for isotropic spherical co-ordinates.
As expected, we retain the zeroth-order part of the original Hamiltonian, while generating the first-order mass
perturbation terms proportional to $\bt$, which couple the right- and left-handed states of the four-spinor.
In addition, we have higher-order mass corrections to the diagonal parts of the Hamiltonian, including a contribution
due to $\Rvec$.
For neutrino interactions, we neglect the higher-order mass terms in (\ref{HCT-Sch-Ortho}) and (\ref{HCT-Sch-Iso}).

\section{High-Energy Helicity and Chirality Transitions in Schwarzschild Space-Time}

In order to determine the high-energy form of the helicity and chirality transition rates, we need to first compute
the helicity and projection operators in the high-energy approximation from (\ref{Xexpand2=}), since
\be
\dot{X}_{\ct} & = & {i \over \hb} \lt[H_{\ct}, X_{\ct}\rt] \ = \ e^{i S_{\rm CT}} \, {i \over \hb} \lt[H, X\rt] e^{-i S_{\rm CT}}
\nn
& = & e^{i S_{\rm CT}} \, \dot{X} \, e^{-i S_{\rm CT}} \, .
\label{Xdot-CT=}
\ee
Then it follows that to first-order in $\omega(q)$,
\be
h_{\ct} & \approx & h \, ,
\label{h-CT=}
\nl
P_{\ct \, \pm} & \approx & P_\pm \mp {1 \over 2} \, {\omega(q) \over p} \, \bt \dotprod{\sg}{p}
 \ = \ P_\pm \mp {1 \over 2} \, {\omega(q) \over p} \, \bt \, h \, .
\label{P-CT=}
\ee
If we write the high-energy Hamiltonians (\ref{HCT-Sch-Ortho}) and (\ref{HCT-Sch-Iso})
in the form $H_{\ct} = H_{\ct}^{(0)} + {\omega(q) \over p} \, \bt \, {\cal H}_{\ct}^{(1)} \,$, then
\be
\dot{h}_{\ct} & \approx & \lt. \dot{h}\rt|_{m = 0} + {\omega(q) \over p} \, \bt \, {i \over \hb} [{\cal H}_{\ct}^{(1)}, h] \, ,
\label{h1dot=}
\nl
\dot{P}_{\ct \, \pm} & \approx &
\pm \, {\omega(q) \over p} \, \bt \lt[{i \over \hb} \, \gfive \, {\cal H}_{\ct}^{(1)} + {i \over \hb} \, H_{\ct}^{(0)} \, h
-  \lt. \dot{h}\rt|_{m = 0} \rt] \, .
\label{P1dot=}
\ee
It is interesting to note from (\ref{P-CT=}) and (\ref{P1dot=}) that the chirality transition rate in the high-energy approximation
is dependent on the zeroth-order contribution of the helicity transition rate, the implications of which are
unclear at present.

By straightforward application of (\ref{Xdot-CT=}) with (\ref{h-CT=}), we show that the leading order expressions for the
high-energy helicity transition rate for a neutrino in Schwarzschild space-time, as seen by a distant observer, are
\be
\dot{h}_{\ct} & \approx & \lt. \dot{h} \rt|_{m = 0}
+ \qp \lt( F \pt{r} \ln F \rt) \lt(1 - {\hb \over p^2} \dotprod{\sg}{R} \rt) \bt
\lt[\lt(2 \, F - 1\rt)\lt(\sgvec^r \pvec^r\rt)\pvec^r + \dotprod{\sg}{p}\pvec^r - \sgvec^r\dotprod{p}{p} \rt]
\label{hdot-CT-Sch-Ortho}
\ee
in Schwarzschild co-ordinates, and
\be
\dot{h}_{\ct} & \approx & \lt. \dot{h} \rt|_{m = 0}
+ \qp \, {R_- \over R_+^3} \lt(1 - {\hb \over p^2} \dotprod{\sg}{R} \rt) \bt
\lt[\pt{\rbar} \ln \lt({R_- \over R_+^3}\rt) \dotprod{\sg}{p} \pvec^{\rbar}
- R_+^2 \pt{\rbar} \ln \lt({R_- \over R_+}\rt) \sgvec^{\rbar}\dotprod{p}{p}
\rt]
\label{hdot-CT-Sch-Iso}
\ee
in isotropic spherical co-ordinates.
In both (\ref{hdot-CT-Sch-Ortho}) and (\ref{hdot-CT-Sch-Iso}), we notice that the high-energy corrections
to the zeroth-order helicity transition rate are dependent on the helicity operator and projected in the
radial direction.
As well, these corrections reduce to zero in the limit as $M \rightarrow 0$, as we should expect.
For a neutrino near the black hole's event horizon, we find that the high-energy correction for
(\ref{hdot-CT-Sch-Ortho}) also grows singular due to gravitational blueshift, while for (\ref{hdot-CT-Sch-Iso})
it is shown that
\be
\lt. \dot{h}_{\ct} \rt|_{\rbar \approx M/2} & \approx &
- {1 \over 4M} \lt[1 - \qp \, \bt \dotprod{\sg}{p}\rt] \pvec^{\rbar}
- {1 \over M} \, \qp \, \bt \, \sgvec^{\rbar}\dotprod{p}{p}
- {\hb \over 4M^2} \, \cot \th \, \alvec^\ph \, .
\label{hdot-CT-Iso-r=M/2}
\ee

Similar application of (\ref{Xdot-CT=}) with (\ref{P-CT=}) shows that the leading-order Hermitian contributions
to the high-energy chirality transition rate in the high-energy approximation are
\be
\dot{P}_{\ct \, \pm} & \approx & \pm \, \qp \, F \, \bt
\lt[-{1 \over \hb} \lt(F - 1\rt)\cross{\al}{p}^r \, \pvec^r + F \lt(\pt{r} \, \ln F^2\rt)\pvec^r
+ \gfive \lt[{F \over r} \lt(1 + r \pt{r} \ln F^{1/2}\rt)\pvec^r
+ {1 \over 2 \, r} \, \cot \th \, \pvec^\th \rt] \rt]
\nn
\label{projdot-CT-Sch-Ortho}
\ee
and
\be
\dot{P}_{\ct \, \pm} & \approx & \pm \, \qp \, {R_- \over R_+^3} \, \bt
\lt[\pt{\rbar} \, \ln \lt({R_- \over R_+^3}\rt)\pvec^{\rbar}
+ \gfive \lt[{1 \over \rbar} \lt(1 + \rbar \pt{\rbar} \ln \lt(R_-R_+^3\rt)^{1/2}\rt) \pvec^{\rbar}
+ {1 \over 2 \, \rbar} \, \cot \th \, \pvec^\th \rt] \rt]
\label{projdot-CT-Sch-Iso}
\ee
in Schwarzschild and isotropic co-ordinates, respectively.
Proceeding similarly as with the helicity transition rate, it is straightforward to show for both
(\ref{projdot-CT-Sch-Ortho}) and (\ref{projdot-CT-Sch-Iso}) that
\be
\lt. \dot{P}_{\ct \, \pm} \rt|_{M = 0} & = & \pm \, \qp \, \bt \, \gfive
\lt[{1 \over r} \, \pvec^r
+ {1 \over 2 \, r} \, \cot \th \, \pvec^\th \rt]
\label{projdot-CT-M=0}
\ee
in the limit $M \rightarrow 0$.
As for neutrino propagation near the black hole's event horizon, it is shown that
\be
\lt. \dot{P}_{\ct \, \pm} \rt|_{r \approx 2M} & \approx & \pm \, {1 \over 8M} \, \qp \, \bt \lt(\gfive + 4\rt)\pvec^r
\label{pdot-CT-Sch-r=2M}
\ee
for Schwarzschild co-ordinates, while
\be
\lt. \dot{P}_{\ct \, \pm} \rt|_{\rbar \approx M/2} & \approx & \pm \, {1 \over 8M} \, \qp \, \bt \lt(\gfive + 2\rt)\pvec^{\rbar}
\label{pdot-CT-Iso-r=M/2}
\ee
for isotropic spherical co-ordinates.
Apart from a factor of two in the second terms of (\ref{pdot-CT-Sch-r=2M}) and (\ref{pdot-CT-Iso-r=M/2}), the chirality
transition rates at the event horizon are identical.

There is one difficulty, however, worth mentioning as it concerns isotropic co-ordinates near the event horizon, in that
the radial momentum operator becomes degenerate when expressed in terms of the Schwarzschild radial co-ordinate, as shown in
(\ref{Pisotropic=}).
This makes the interpretation of (\ref{hdot-CT-Iso-r=M/2}) and (\ref{pdot-CT-Iso-r=M/2}) more difficult because the
co-ordinate system is ill-defined at the event horizon, once the $\rbar$-dependence is rewritten in terms of $r$.

\section{Special Considerations}

\subsection{Weak-Field Limit with Gravitational Berry's Phase}

In order to identify neutrino helicity and chirality interactions in an observationally realistic
context, it is necessary to consider the weak-field limit of curved space-time.
To do this requires that we perform a first-order expansion about Minkowski space-time,
such that the metric is described by $\vec{g} = (\eta_{\alpha \beta} + h_{\alpha \beta}) \,
\d x^\alpha \otimes \d x^\beta$, and the covariant Dirac equation (\ref{covDirac=}) takes an
equivalent form for a local tangent space centred about the gravitational source.
An interesting development proposed by Cai and Papini \cite{Cai} is that the phase of the
wavefunction can contain gravitational field information that leads to a covariant
generalization of Berry's phase \cite{Berry} when considering closed particle trajectories,
with an explicit expression for the case of the weak-field limit.
It is further shown \cite{Singh} that the gravitational phase can be introduced into the
Dirac Hamiltonian as a vector gauge field within the framework of Minkowski space-time.

While a more complete treatment of the gravitational Berry's phase is found in
Appendix~\ref{appendix:Berry},
it suffices to state that (\ref{covDirac=}) is equivalent to
\be
\lt[i \gamma^{\hat{\mu}} \lt(\nabla_{\hat{\mu}} + i \Gamma_{\hat{\mu}}
+ \frac{i}{\hbar} (\nabla_{\hat{\mu}} \Phi_{\rm G}) \rt) -
{m \over \hbar} \rt]\psi (x) & \approx & 0 \, ,
\ee
where $\Phi_{\rm G}$, the gravitational Berry's phase, takes on a non-zero value only for
{\em closed} trajectories in space-time \cite{Cai}.
Because $\Phi_{\rm G}$ was originally derived in terms of a Cartesian co-ordinate frame,
it is best to consider the weak-field limit of this problem in the same fashion.
Therefore, given the one-form basis set corresponding to the weak-field limit in Cartesian co-ordinates,
%
%
%
\be
\vec{e}^{\hat{0}} & = & \lt(1 - {2M \over r}\rt)^{1/2} \, \d t \, ,
\nn
\vec{e}^{\hat{\imath}} & = & \lt(1 + {2M \over r}\rt)^{1/2} \, \d x^{\hat{\imath}} \, ,
\label{1-form-basis=}
\ee
where $(x^{\hat{1}}, x^{\hat{2}}, x^{\hat{3}}) = \lt(x, y, z\rt)$
and $r = \sqrt{x^2 + y^2 + z^2}$, a straightforward calculation shows that the Dirac Hamiltonian up to
first-order in $M/r$ is
\be
H & \approx &
m \lt(1 - {M \over r}\rt) \beta
+ \lt(1 - {2M \over r}\rt)\dotprod{\al}{p} + {\ihb \over 2} \, {M \over r^2} \lt(\alvec \cdot \nabvec r\rt)
+ \lt(\alvec \cdot \nabvec \PhiG \rt) + \lt(\nabla_0 \PhiG \rt) \, ,
\label{H-WF-cart=}
\ee
where $\pvec^{\hat{\imath}} = -\ihb \, \partial / \partial x^{\hat{\imath}}$.
While it is clear from Appendix~\ref{appendix:Berry} that the gravitational
Berry's phase is Lorentz invariant, it is important to emphasize here that the gradients of $\Phi_{\rm G}$ are
{\em frame-dependent} and will vary in appearance for different co-ordinate systems.
Applying the CT transformation to (\ref{H-WF-cart=}), we find that
\be
H_{\ct} & \approx & \lt(1 - {2M \over r}\rt)\dotprod{\al}{p} +  {\ihb \over 2} \, {M \over r^2} \lt(\alvec \cdot \nabvec r\rt)
+ \lt(\alvec \cdot \nabvec \PhiG \rt) + \lt(\nabla_0 \PhiG \rt)
\nn
& &{} + \qp \, \bt \lt[{M \over r} \dotprod{p}{p} - \hb \, {M \over r^2} \, \sgvec \cdot \lt(\nabvec r \times \pvec\rt) \rt]
+ {1 \over 2} \, q^2 \lt(1 - {M \over r}\rt)\dotprod{\al}{p} \, ,
\label{HCT-Sch-WF}
\ee
where the second term in the linear mass-perturbation contribution to (\ref{HCT-Sch-WF}) indicates a
spin-orbital coupling interaction present.
It follows that the helicity transition rate for a high-energy neutrino in the weak-field limit of
Schwarzschild space-time is
\be
\dot{h}_{\ct} & \approx & \lt. \dot{h} \rt|_{m = 0}
+ \qp \, \bt \,
\lt[{2M \over r}\lt(\nabvec r \cdot \pvec\rt)\dotprod{\sg}{p} - {M \over r}\lt(\sgvec \cdot \nabvec r\rt)\dotprod{p}{p}\rt] \, ,
\label{hdot-CT-Sch-WF}
\ee
where
\be
\dot{h} & = & -m \, {M \over r^2} \, \bt \lt(\sgvec \cdot \nabvec r\rt) - {2M \over r^2} \lt(\nabvec r \cdot \pvec\rt)
- {2 \over \hb} \, \alvec \cdot \lt(\nabvec \PhiG \times \pvec\rt) - \gfive \, \nabvec^2 \PhiG - \sgvec \cdot \nabvec \lt(\nabla_0 \PhiG \rt)
\nn
& &{} - 3 \, i \, {M \over r^2} \, \alvec \cdot \lt(\nabvec r \times \pvec\rt)
+ \ihb \, \gfive \, {M \over r^2} \lt[{1 \over r} \lt(\nabvec r \cdot \nabvec r\rt) - {1 \over 2} \, \nabvec^2 r\rt] \, .
\label{hdot-Sch-WF}
\ee
As well, the weak-field approximation of the chirality transition rate for a high-energy neutrino is
\be
\dot{P}_{\ct \, \pm} & \approx & \pm \qp \, \bt
\lt[{2M \over r^2}\lt(1 - {1 \over 4} \, \gfive\rt)\lt(\nabvec r \cdot \pvec\rt)
+ {2 \over \hb} \, \alvec \cdot \lt(\nabvec \PhiG \times \pvec \rt)
+ \gfive \, \nabvec^2 \PhiG + \sgvec \cdot \nabvec \lt(\nabla_0 \PhiG\rt) \rt] \, .
\label{projdot-CT-Sch-WF}
\ee

For comparison, we can derive an expression equivalent to (\ref{H-WF-cart=}) in spherical
co-ordinates by performing a first-order expansion of $M/r$ in
(\ref{H-Schw-isotropic=}).  Then the Dirac Hamiltonian becomes
\be
H & \approx &
m \lt(1 - {M \over r}\rt) \beta
+ \lt(1 - {2M \over r}\rt)\dotprod{\al}{p} + {\ihb \over r} \lt(1 - {5M \over 2 \, r} \rt) \alvec^r
- {\ihb \over 2 \, r} \, \lt(1 - {2M \over r}\rt) \cot \th \, \alvec^\th
\nn
& &{} + \lt(\alvec \cdot \nabvec \PhiG \rt) + \lt(\nabla_0 \PhiG \rt)
\label{H-WF-sph=}
\ee
in the weak-field limit, where $\vec{p}$ is described by (\ref{momentum1=}).
It is trivial to note--but important to emphasize--that the free-particle
Hamiltonian expressions from (\ref{H-WF-cart=}) and (\ref{H-WF-sph=}) are distinct
in the limit as $M / r \rightarrow 0$, due to the choice of co-ordinate system.
Again, by applying the CT transformation, we obtain for the high-energy approximation
of the Hamiltonian
\be
H_{\ct} & \approx & \lt(1 - {2M \over r}\rt)\dotprod{\al}{p} + {\ihb \over r} \lt(1 - {5M \over 2 \, r} \rt) \alvec^r
- {\ihb \over 2 \, r} \lt(1 - {2M \over r}\rt) \cot \th \, \alvec^\th + \lt(\alvec \cdot \nabvec \PhiG \rt) + \lt(\nabla_0 \PhiG \rt)
\nn
& &{} + \qp \, \bt \lt[{M \over r} \lt[\dotprod{p}{p} + \hb \dotprod{\sg}{R}\rt]
+ \hb \, {M \over r} \cross{\sg}{p}^r \rt]
+ {1 \over 2} \, q^2 \lt(1 - {M \over r}\rt)\dotprod{\al}{p} \, ,
\label{HCT-Sch-WF-Sph}
\ee
where a similar type of spin precession term appears in the linear mass-perturbation correction.
Then the high-energy helicity transition rate follows as
\be
\dot{h}_{\ct} & \approx & \lt. \dot{h} \rt|_{m = 0}
+ \qp \, \bt
\lt[{2M \over r}\dotprod{\sg}{p}\pvec^r - {M \over r} \, \sgvec^r \dotprod{p}{p}\rt] \, ,
\label{hdot-CT-Sch-WF-Sph}
\ee
where
\be
\dot{h} & = & -m \, {M \over r^2} \, \bt \, \sgvec^r - {2M \over r^2} \, \pvec^r
+ {\hb \over 2 \, r^2} \lt(1 - {4M \over r}\rt) \cot \th \, \alvec^\ph
- {2 \over \hb} \, \alvec \cdot \lt(\nabvec \PhiG \times \pvec\rt) - \gfive \, \lt(\nabvec \cdot \nabvec \PhiG\rt)
- \sgvec \cdot \nabvec \lt(\nabla_0 \PhiG \rt)
\nn
& &{} - {i \over r} \lt[2 \lt(1 - {7M \over 2 \, r}\rt)\cross{\al}{p}^r + \lt(1 - {2M \over r}\rt) \cot \th \, \cross{\al}{p}^\th \rt]
- i \cross{\al}{\nabla} \cdot \nabvec \PhiG
\nn
& &{} - {\ihb \over r^2} \, \gfive \lt[\lt(1 - {5M \over r}\rt) + {1 \over 2} \, {1 \over \sin^2 \th} \lt(1 - {2M \over r}\rt)\rt] \, ,
\label{hdot-Sch-WF-Sph}
\ee
and $\nabvec \cdot \nabvec = \nabvec_{\hat{1}} \nabvec_{\hat{1}} + \nabvec_{\hat{2}} \nabvec_{\hat{2}} + \nabvec_{\hat{3}} \nabvec_{\hat{3}} \,$,
which is only equal to the Laplacian operator in Cartesian co-ordinates.
Similarly, the expression for the chirality transition rate of a high-energy neutrino is
\be
\dot{P}_{\ct \, \pm}  & \approx & \pm \, \qp \, \bt
\lt[{2M \over r^2} \, \pvec^r
- \gfive
\lt[{1 \over r} \lt(1 - {5M \over 2 \, r} \rt)\pvec^r - {1 \over 2 \, r} \lt(1 - {2M \over r} \rt)
\cot \th \, \pvec^\th \rt]\rt.
\nn
& &{} + \lt. {2 \over \hb} \, \alvec \cdot \lt(\nabvec \PhiG \times \pvec \rt)
+ \gfive \lt(\nabvec \cdot \nabvec \PhiG \rt) + \sgvec \cdot \nabvec \lt(\nabla_0 \PhiG\rt)\rt] \, .
\nn
\label{projdot-CT-Sch-WF-Sph}
\ee

A comparison between the two sets of calculations in the weak-field limit suggests that, in the limit as
$M / r \rightarrow 0$, the helicity defined in Cartesian co-ordinates is a constant of the motion, while the same
is apparently not true in spherical co-ordinates.
However, this is merely a co-ordinate effect that comes from the spinor connection in Minkowski space-time.
To demonstrate this more precisely, we can calculate the expectation value of the helicity transition rate
for a local observer comoving with the neutrino, such that
\be
{dh \over d\tau} & = & \lt[g_{00}(x)\rt]^{-1/2} \, \dot{h}
\label{hdot-tau}
\ee
for proper time $\tau$.
We can define a normalized state vector
$\lt| \Psi_\pm \rt\rangle \equiv \lt| \pm \rt\rangle \otimes \lt| \psi_{\rm C} \rt\rangle \, $,
where $\lt| \psi_{\rm C} \rt\rangle$ is described by (\ref{psi-chiral}),
and
\be
\lt| + \rt\rangle & = & \left(
\begin{array}{c}
1 \\
0
\end{array} \right), \qquad
\lt| - \rt\rangle \ = \ \left(
\begin{array}{c}
0 \\
1
\end{array} \right) \, ,
\label{+-spinors}
\ee
are the neutrino's positive and negative helicity states with respect to its quantization axis
in the radial direction, such that
\be
\langle \mp | \vec{\sigma} | \pm \rangle & = &
\left[\cos \theta \, \cos \varphi \mp i \, \sin \varphi \right] \vec{\hat{x}} +
\left[\cos \theta \, \sin \varphi \pm i \, \cos \varphi \right] \vec{\hat{y}} -
\sin \theta \, \vec{\hat{z}}
\label{orientation=}
\ee
for Cartesian co-ordinates, and
\be
\lt\langle \mp \rt| \sgvec \lt| \pm \rt\rangle & = & \vec{\hat{x}}^\th \pm i \, \vec{\hat{x}}^\ph
\label{orientation2=}
\ee
for spherical co-ordinates.
Then it can be shown that the high-energy helicity transition rate is
$\lt\langle dh / d\tau \rt\rangle_{\ct \, \pm}
 \equiv \ \lt\langle \Psi_\mp \rt| \lt(dh / d\tau\rt)_{\ct} \lt| \Psi_\pm \rt\rangle \,$,
where for Cartesian co-ordinates it follows from (\ref{hdot-CT-Sch-WF}) and (\ref{hdot-Sch-WF}) that
\be
\lt\langle {dh \over d\tau} \rt\rangle_{\ct \, \pm} & \approx &
-\lt\langle \mp \rt| \sgvec \lt| \pm \rt\rangle \cdot \lt[\nabvec \lt(\nabla_0 \PhiG\rt) +
\lt(\lt|C_{\rm R}\rt|^2 - \lt|C_{\rm L}\rt|^2\rt) \lt[{2 \over \hb} \, \lt(\nabvec \PhiG \times \pvec \rt)
+ 3 \, i \, {M \over r^2} \lt(\nabvec r \times \pvec\rt) \rt] \rt.
\nn
& &{} + \lt. 2 \, {\rm Re} \lt(C_{\rm R} \, C_{\rm L}^*\rt) \qp \lt[{2M \over r} \lt(\nabvec r \cdot \pvec\rt)\pvec
- {M \over r} \, \nabvec r \dotprod{p}{p}\rt] \rt] \, .
\label{<dh-dtau>-Cart}
\ee
It is obvious here that (\ref{<dh-dtau>-Cart}) vanishes when $M / r \rightarrow 0$.
Performing the same analysis for the spherical co-ordinate frame, we can show from (\ref{hdot-CT-Sch-WF-Sph}) and (\ref{hdot-Sch-WF-Sph})
that
\be
\lt\langle {dh \over d\tau} \rt\rangle_{\ct \, \pm} & \approx &
-\lt\langle \mp \rt| \sgvec \lt| \pm \rt\rangle \cdot \lt[\nabvec \lt(\nabla_0 \PhiG\rt) +
\lt(\lt|C_{\rm R}\rt|^2 - \lt|C_{\rm L}\rt|^2\rt) \lt[{2 \over \hb} \, \lt(\nabvec \PhiG \times \pvec \rt)
+ \hb \lt(\Rvec \, \PhiG\rt) \rt] \rt]
\nn
& &{} \mp \lt(\lt|C_{\rm R}\rt|^2 - \lt|C_{\rm L}\rt|^2\rt) \lt[{2 \over r} \lt(1 - {5M \over 2r}\rt) \lt(\pvec^\th \pm i \, \pvec^\ph\rt)
- {1 \over r} \lt(1 - {M \over r}\rt) \cot \th \, \pvec^r - {\ihb \over 2 \, r^2} \lt(1 - {3M \over r}\rt) \cot \th \rt]
\nn
& &{} - 2 \, {\rm Re} \lt(C_{\rm R} \, C_{\rm L}^* \rt) \qp \, {2M \over r} \lt(\pvec^\th \pm i \, \pvec^\ph\rt)\pvec^r \, ,
\label{<dh-dtau>-Sph}
\ee
which is generally non-zero.
However, without any loss of generality it is evident that,
for exclusively radial propagation $\lt(\pvec^\th = \pvec^\ph = 0\rt)$ along the
$\th = \pi/2$ plane, it follows that (\ref{<dh-dtau>-Sph}) vanishes as well in the $M / r \rightarrow 0$ limit.
Similarly, it is straightforward to show that the helicity-flip induced chirality transition rates for
(\ref{projdot-CT-Sch-WF}) and  (\ref{projdot-CT-Sch-WF-Sph}) are
\be
\lt\langle \dot{P}_{\ct \, \pm} \rt\rangle_\pm & \approx & \mp \, 2 \, {\rm Re} \lt(C_{\rm R} \, C_{\rm L}^* \rt)
\qp \lt(\vec{\hat{x}} \pm i \, \vec{\hat{y}} \rt) \cdot \nabvec \lt(\nabla_0 \PhiG\rt)
\label{<projdot>-Cartesian}
\ee
in Cartesian co-ordinates, and
\be
\lt\langle \dot{P}_{\ct \, \pm} \rt\rangle_\pm
& \approx & \mp \, 2 \, {\rm Re} \lt(C_{\rm R} \, C_{\rm L}^* \rt)
\qp \lt(\vec{\hat{x}}^\th \pm i \, \vec{\hat{x}}^\ph \rt) \cdot \nabvec \lt(\nabla_0 \PhiG\rt)
\label{<projdot>-spherical}
\ee
in spherical co-ordinates.

A comparison of (\ref{<dh-dtau>-Cart})--(\ref{<projdot>-spherical}) shows that the gravitational Berry's phase
makes a meaningful contribution to the high-energy helicity and chirality transition rates.
In particular, the term $\nabvec \lt(\nabla_0 \PhiG\rt)$ persists regardless of the neutrino's degree of handedness
or mass dependence where it concerns the helicity transition rate.
One intriguing possibility that arises here is the potential ability to determine the absolute mass of a neutrino
and the amount of right-handedness within a typical neutrino beam from measuring the helicity and chirality
transition rates, and performing a parameter fit on the data to determine $C_{\rm R}\,$, $C_{\rm L}\,$, and $q \,$.
In practice, however, such experiments would be extremely difficult to perform because of the need for an enormous
amount of neutrino flux to get statistically significant measurements.
According to the Standard Model \cite{Mohapatra}, neutrinos have only negative helicity and are exclusively left-handed.
Given the Superkamiokande and SNO evidence in favour of neutrinos with rest masses, we should expect some right-handed
states to exist (at least for Dirac neutrinos), but to the degree that $\lt|C_{\rm R}\rt| \ll 1$ and $\lt|C_{\rm L}\rt| \sim 1$.
Unless this assumption is shown to be completely wrong, there will be practically no linear mass contribution to the
neutrino helicity transition rate and effectively no observable chirality transition rate whatsoever.
Therefore, it would be a great surprise if we were to obtain an unambiguously non-zero measurement of the helicity and
chirality transition rates with discernable mass dependence, based on (\ref{<dh-dtau>-Cart})--(\ref{<projdot>-spherical}).

\subsection{Chiral Anomaly}

Although this paper is primarily interested in the quantum {\em mechanics} of neutrinos in curved space-time, there are
legitimate reasons to briefly consider potential implications to quantum field theory in this context.
It is beyond the scope of this paper to consider a detailed treatment of neutrino quantum field theory in a gravitational
background, but one particularly relevant issue to examine is the possible impact of the CT transformation on the
chiral anomaly \cite{Holstein} in curved space-time.
It is understood \cite{Mielke} that the conservation of the chiral current in the massless limit is violated by the
Pontrjagin topological invariant, which is proportional to the square of the Riemann curvature tensor.
Here, we examine whether the CT transformation has any significant impact on the known expression
and discuss its possible implications.

To begin, we have the action for the Dirac field
\be
S & = & \int d^4 x \, \sqrt{-g} \, {\cal L}_{\rm D} \, ,
\label{action}
\ee
where the Lagrangian density ${\cal L}_{\rm D}$ is
\be
{\cal L}_{\rm D} & = & \bar{\psi}(x) \lt[i \gm^\mu(x) \, {\cal D}_\mu - {m \over \hbar} \rt] \psi(x) \, ,
\label{Lagrangian}
\ee
and $\bar{\psi}(x) = \psi^\dag(x) \, \bt$.
We know that if we perform an infinitesimal chiral transformation \cite{Ramond}
\be
\psi(x) \rightarrow \exp\lt[i \epsilon(x) \, \gfive\rt]\psi(x) \approx \lt[1 + i \epsilon(x) \, \gfive\rt] \psi(x) \, ,
\label{chiral-trans}
\ee
the chiral current $J_5^\mu = i \sqrt{-g} \, \bar{\psi} \, \gfive \, \gm^\mu(x) \, \psi$ satisfies the equation
\be
0 & = & {1 \over \sqrt{-g}} \, \partial_\mu J_5^\mu
- {2 \, m \over \hb} \, \bar{\psi} \, \gfive \, \psi \, .
\label{gamma5-eq}
\ee
If we now perform the CT transformation on (\ref{gamma5-eq}), we obtain to leading order in $q$
\be
0 & \approx & {1 \over \sqrt{-g}} \, \partial_\mu J_{5 \, \ct}^\mu
- {2 \, m \over \hb} \, \bar{\psi}_{\rm CT} \, \gfive \, \psi_{\rm CT}
\nn
& &{} - \qp \, {1 \over \sqrt{-g}} \, \partial_\mu \lt[\sqrt{-g} \, \bar{\psi}_{\rm CT} \, \gfive
\lt[{\hb \over 2} \, \bt \lt[\dotprod{\al}{\nabla} \gm^\mu\rt] + e^\mu{}_{\hat{\jmath}} \cross{\sg}{p}^{\hat{\jmath}}
- i \, e^\mu{}_{\hat{0}} \dotprod{\al}{p} \rt]\psi_{\rm CT} \rt] \, ,
\label{g5-eq-CT}
\ee
where
\be
J_{5 \, \ct}^\mu & = & i \sqrt{-g} \, \bar{\psi}_{\rm CT} \, \gfive \, \gm^\mu(x) \, \psi_{\rm CT} \, .
\label{j5-CT}
\ee
We see from (\ref{g5-eq-CT}) that the mass-dependent correction to the chiral current is frame-dependent, due
to the presence of $\pvec$ in its definition.
Clearly, when taking the massless limit, we obtain the conservation of chiral current which follows from (\ref{gamma5-eq}).
Therefore, it follows that while there is a small mass correction to the original chiral anomaly equation \cite{Mielke},
we expect no impact on the chiral anomaly in the massless limit, since the Pontrjagin invariant comes from the
spinor connection $\Gamma_\mu$, which is not present in (\ref{g5-eq-CT}).

It should be noted that there is an ambiguity in determining the modified chiral current equation, because if we instead
performed the CT transformation first, followed by the chiral transformation, we obtain in place of (\ref{g5-eq-CT})
\be
0 & \approx & {1 \over \sqrt{-g}} \, \partial_\mu J_{5 \, \ct}^\mu
- {2 \, m \over \hb} \, \bar{\psi}_{\rm CT} \, \gfive \, \psi_{\rm CT}
\nn
& &{} - \qp \lt\{ {1 \over \sqrt{-g}} \, \partial_\mu \lt[\sqrt{-g} \, \bar{\psi}_{\rm CT} \, \gfive
\lt[{\hb \over 2} \, \bt \lt[\dotprod{\al}{\nabla} \gm^\mu\rt] - i \, e^\mu{}_{\hat{\jmath}} \, \pvec^{\hat{\jmath}} \rt]\psi_{\rm CT} \rt]
\rt.
\nn
& &{} - \lt. \bar{\psi}_{\rm CT} \, \gfive
\lt[{\hb \over 2} \, \bt \lt[\dotprod{\al}{\nabla} \gm^\al\rt] - i \, e^\al{}_{\hat{\jmath}} \, \pvec^{\hat{\jmath}} \rt]\partial_{\al}\psi_{\rm CT}
-i \, \bar{\psi}_{\rm CT} \, \gfive
\lt[
{\hb \over 2} \, \bt \lt[\dotprod{\al}{\nabla} \gm^\al \, \Gamma_\al\rt]
+ {i \over 2} \lt\{\bt \, \alvec, \gm^\al \, \Gamma_\al\rt\} \cdot \pvec \rt]\psi_{\rm CT}
\rt\} \, ,
\nn
\label{g5-eq-CT-2}
\ee
which has a non-trivial dependence on $\Gamma_\mu \,$.
However, it seems more correct to follow the approach leading to (\ref{g5-eq-CT}) because the original chiral current
equation (\ref{gamma5-eq}) follows from an {\em exact} Dirac Lagrangian density, while the chiral transformation
leading to (\ref{g5-eq-CT-2}) comes from an {\em approximation} of ${\cal L}_{\rm D}$, so the correspondence is not exact.

\section{Conclusion}

In this paper, we determined the helicity and chirality transition rates for a high-energy massive neutrino
propagating in a Schwarzschild space-time background, where the dynamics is governed by a
Hamiltonian transformed by the Cini-Touschek unitary operator.
We derived the high-energy Hamiltonian to first order in $q = m/p$ in terms of both regular Schwarzschild
co-ordinates and isotropic spherical co-ordinates, and determined the helicity and chirality
transition rates accordingly, with particular attention paid to physical effects near the event horizon
of a black hole.
Under general conditions, we learned that the neutrino's helicity transition rate in a gravitational
background is not a constant of the motion for massless particles, while the chirality transition rate still
retains an overall dependence on neutrino mass, and thus vanishes in the massless limit.
It was demonstrated that a relationship exists between the chiral transition rate and the
zeroth-order contribution of the helicity transition rate in the high-energy approximation.
We further considered the special case of a weak gravitational field with the inclusion of
the gravitational analogue of Berry's phase, where the expectation values for the helicity and
chirality transition rates were obtained and compared.
In addition, we determined the mass-dependent corrections to the chiral current equation and its
relationship to the chiral anomaly when making use of the Cini-Touschek transformation.

For this investigation, we used the helicity operator as defined in flat space-time.
Strictly speaking, the more appropriate operator to use is the time component of the Pauli-Lubanski
four-vector \cite{Itzykson} defined in curved space-time.
The next development for this research is to compute the covariant spin dynamics of
neutrinos in Schwarzschild and Kerr backgrounds using the Pauli-Lubanski vector \cite{Singh2},
which can be compared to the results contained in this paper.
Cosmological considerations during the earliest stages of the known Universe may be relevant
for applications of this type of research.
Another possibility of future development is to consider a detailed study of neutrino
wave packets within this context, to see how the spin and chiral behaviour would differ
for a wave packet description, as opposed to plane waves.
These and other types of possibilities may be developed in the near future.

\section{Acknowledgements}

We are grateful to Nader Mobed at the University of Regina for his encouragement and
insightful comments throughout the completion of this paper.
As well, we thank Bahram Mashhoon and Giorgio Papini for many valuable correspondences
and suggestions.
We also thank the anonymous referee for suggestions given to improve this paper, particularly
the recommendation to investigate the chiral anomaly problem.

\begin{appendix}

\section{Gravitational Berry's Phase in the Weak-Field Limit}
\label{appendix:Berry}
\renewcommand{\theequation}{A.\arabic{equation}}
\setcounter{subsection}{0}
\setcounter{equation}{0}

Berry's phase \cite{Berry} was originally formulated in terms of the non-relativistic
Schr\"{o}dinger equation, assuming adiabatic transport of the quantum system around a
closed path in parameter space.
It is a geometric and gauge-invariant phase that is identified with the motion of the
system in the space of parameters.
One generalization of this formalism \cite{Dandoloff} is to evolve the system using
Fermi-Walker transport in place of parallel transport, which has the advantage of
applicability for non-geodesic motion.

The covariant generalization of Berry's phase, as developed by Cai and Papini \cite{Cai},
allows for the introduction of Lorentz-invariant classical fields within a quantum
framework.
One example of this is the covariant generalization of the Aharonov-Bohm effect \cite{Cai,Aharonov}.
In this Appendix, we present a brief description of the gravitational
Berry's phase for the weak-field limit of Schwarzschild space-time.
This is formally described by the line integral
\be
\Phi_{\rm G} & \equiv & \frac{1}{2} \int_{x_0}^x dz^{\hat{\lambda}} \,
h_{\hat{\alpha} \hat{\lambda}} (z) \pvec^{\hat{\alpha}}
 + \frac{1}{4} \int_{x_0}^x dz^{\hat{\lambda}} \lt[h_{\hat{\beta} \hat{\lambda},
\hat{\alpha}} (z) - h_{\hat{\alpha} \hat{\lambda}, \hat{\beta}} (z) \rt]
\vec{L}^{\hat{\alpha} \hat{\beta}} (z) \, ,
\label{Berry=}
\ee
where the integration path is specified along the geodesic that describes
the particle's motion,
$\pvec^{\hat{\mu}}$ is the momentum of a free particle described by the wavefunction
$\Psi_0$ and $\vec{L}^{\hat{\alpha} \hat{\beta}}$ is its corresponding orbital angular momentum
satisfying
\be
\lt[\vec{L}^{\hat{\alpha} \hat{\beta}} (z), \Psi_0 \rt] & \equiv &
\lt[ (x^{\hat{\alpha}} - z^{\hat{\alpha}}) \pvec^{\hat{\beta}} -
(x^{\hat{\beta}} - z^{\hat{\beta}}) \pvec^{\hat{\alpha}} \rt]
\Psi_0 \, ,
\label{L=}
\nl
\lt[\pvec^{\hat{\alpha}} , \Psi_0 \rt] & \equiv &  \ihb \, \partial^{\hat{\alpha}} \,
\Psi_0 \, .
\label{P=}
\ee

Formally, it happens that when considering transport of the quantum system in a closed path $C$
in space-time according to (\ref{Berry=}), it can be shown that
the gravitational Berry's phase can be identified with
\be
\PhiG(C)& = & {1 \over 2} \, \oint_C \lt(\Gamma_{\hat{\al}}^{(T)} \, \pvec^{\hat{\al}}
+ \Gamma_{\hat{\al}\hat{\bt}}^{(L)} \, \vec{L}^{\hat{\al}\hat{\bt}} \rt) \, ,
\label{Berry-closed}
\ee
where
\begin{subequations}
\be
\Gamma_{\hat{\al}}^{(T)} & = & dz^{\hat{\lambda}} \, h_{\hat{\alpha} \hat{\lambda}} \, ,
\nl
\Gamma_{\hat{\al}\hat{\bt}}^{(L)} & = & - dz^{\hat{\lambda}} \, \Gamma_{\hat{\al}\hat{\bt}\hat{\lambda}}
\ = \ - dz^{\hat{\lambda}} \lt[
{1 \over 2} \lt(h_{\hat{\al} \hat{\bt}, \hat{\lambda}}
+  h_{\hat{\alpha} \hat{\lambda}, \hat{\beta}} - h_{\hat{\beta} \hat{\lambda}, \hat{\alpha}}\rt) \rt] \, ,
\label{connections}
\ee
\end{subequations}
are the translational and rotational components of the Cartan connection \cite{Morales}.

In Cartesian co-ordinates, it follows that the most general expression for (\ref{Berry=}) is
\be
\Phi_{\rm G} & = & \int_{x_0} ^x \, dt' \, \lt(\nabvec_t \Phi_{\rm G}\rt) +
\int_{x_0} ^x \, dx' \, \lt(\nabvec_x \Phi_{\rm G}\rt) +
\int_{x_0} ^x \, dy' \, \lt(\nabvec_y \Phi_{\rm G}\rt) +
\int_{x_0} ^x \, dz' \, \lt(\nabvec_z \Phi_{\rm G}\rt) \, ,
\label{Berry1a=}
\ee
where
\be
\lt(\nabvec_\mu \Phi_{\rm G}\rt) & = & \lt[{M \over r'^3} \lt( \vec{r} \cdot \vec{r'}\rt) -
{2M \over r'} \rt] \pvec^{\hat{\mu}} - {M \over r'^3} \lt(x^\mu - {x'}^\mu \rt) \vec{r'} \cdot \vec{p} \, .
\ee
By a simple co-ordinate transformation, the gravitational Berry's phase in
spherical co-ordinates is
\be
\Phi_{\rm G} & = & \int_{x_0} ^x \, dt' \, \lt(\nabvec_t \Phi_{\rm G}\rt) +
\int_{x_0} ^x \, dr' \, \lt(\nabvec_r \Phi_{\rm G}\rt) +
\int_{x_0} ^x \, d\th' \, \lt(\nabvec_\th \Phi_{\rm G}\rt) +
\int_{x_0} ^x \, d\ph' \, \lt(\nabvec_\ph \Phi_{\rm G}\rt),
\label{Berry2=}
\ee
where
\be
\lt(\nabvec_t \Phi_{\rm G}\rt) & = & -{2M \over r'} \, \pvec^t +
{M \over r'^2} \lt[\cos\th \, \cos\th' + \sin\th \, \sin\th' \, \cos(\ph - \ph')\rt]
\lt[r \, \pvec^t - (t - t') \, \pvec^r \rt] \, ,
\nl
\lt(\nabvec_r \Phi_{\rm G}\rt) & = & -{M \over r'}
\lt[\cos\th \, \cos\th' + \sin\th \, \sin\th' \, \cos(\ph - \ph')\rt] \pvec^r \, ,
\nl
\lt(\nabvec_\th \Phi_{\rm G}\rt) & = & 2M
\lt[\cos\th \, \sin\th' - \sin\th \, \cos\th' \, \cos(\ph - \ph')\rt]\pvec^r \, ,
\nl
\lt(\nabvec_\ph \Phi_{\rm G}\rt) & = & -2M
\, \sin\th \, \sin\th' \, \sin(\ph - \ph') \, \pvec^r \, ,
\ee
for $\pvec^t = -\ihb \, \partial / \partial t$ and $\pvec^r = -\ihb \, \partial / \partial r$.

\end{appendix}


\begin{thebibliography}{99}

\bibitem{Kam}
Y. Fukuda et al. (Superkamiokande Collaboration), {\it Phys. Rev. Lett.} {\bf 86}, 5656-5660 (2001);
{\bf 86}, 5651-5655 (2001); {\bf 82}, 2644-2648 (1999);
{\it Phys. Lett.} {\bf B433}, 9-18 (1998); {\bf 436}, 33-41 (1998);
{\bf 467}, 185-193 (1999).

\bibitem{SNO}
Q. R. Ahmad et al. (SNO Collaboration) {\it Phys. Rev. Lett.} {\bf 89}, 011302 (2002);
{\bf 89}, 011301 (2002); {\bf 87}, 071301 (2001).


\bibitem{Adak}
M. Adak, T. Dereli, and L. H. Ryder, {\it Class. Quant. Grav.} {\bf 18}, 1503-1512 (2001).

\bibitem{qviolation}
A. Raychaudhuri and A. Sil, {\it Phys. Rev.} {\bf D65}, 073035 (2002);
D. Majumdar, A. Raychaudhuri, and A. Sil, {\bf D63}, 073014 (2001);
H. Casini, J. C. D'Olivo, and R. Montemayor, {\bf D61}, 105004 (2000);
S. Capozziello and G. Lambiase, {\it Gen. Rel. Grav.} {\bf 34}, 1097-1106 (2002).

\bibitem{Minakata}
H. Minakata and H. Nunokawa, {\it Phys. Rev.} {\bf D51}, 6625-6634 (1995).

\bibitem{Choudhury}
D. Choudhury, N. D. Hari Dass, and M. V. N. Murthy, {\it Class. Quant. Grav.} {\bf 6},
L167-L171 (1989).

\bibitem{Zuber}
K. Zuber, {\it Phys. Rep.} {\bf 305}, 295-364 (1998).

\bibitem{Casini}
H. Casini and R. Montemayor, {\it Phys. Rev.} {\bf D50}, 7425-7429 (1994).

\bibitem{Cini}
M. Cini and B. Touschek, {\it Nuovo Cim.} {\bf 7}, 422-424 (1958);
S. K. Bose, A. Gamba, and
E. C. Sudarshan, {\it Phys. Rev.} {\bf 113}, 1661-1663 (1959).

\bibitem{Cai}
Y. Q. Cai and G. Papini, {\it Phys. Rev. Lett.} {\bf 66}, 1259-1262 (1991); {\bf 68}, 3811 (1992);
{\it Class. Quantum Grav.} {\bf 7}, 269-275 (1990);
{\bf 6}, 407-418 (1989);
{\it Gen. Rel. Grav.} {\bf 22}, 259-267 (1990);
{\it Mod. Phys. Lett.} {\bf A4}, 1143-1149 (1990).

\bibitem{Schmutzer}
E. Schmutzer and J. Pleba\'{n}ski, {\it Fortschr. Phys.} {\bf 25}, 37-82 (1977).

\bibitem{Hehl1}
F. W. Hehl, J. Lemke, and E. W. Mielke, ``Two Lectures on Fermions and Gravity,'' in
{\it Geometry and Theoretical Physics}, Eds. J. Debrus and A.C. Hirshfeld, 56-140, Springer-Verlag, Berlin (1991).

\bibitem{Hehl2}
B. Mashhoon, {\it Phys. Rev. Lett.} {\bf 61}, 2639-2642 (1988);
F. W. Hehl and W.-T. Ni, {\it Phys. Rev.} {\bf D42}, 2045-2048 (1990);
Y. N. Obukhov, {\it Fortschr. Phys.} {\bf 50}, 711-716 (2002).

\bibitem{COW}
R. Colella, A. W. Overhauser, and S. A. Werner, {\it Phys. Rev. Lett.} {\bf 34}, 1472-1474 (1975);
S. A. Werner, J.-L. Staudenmann, and R. Colella, {\bf 42}, 1103-1106 (1979);
U. Bonse and T. Wroblewski, {\bf 51}, 1401-1404 (1983).

\bibitem{Mashhoon}
B. Mashhoon, {\it Phys. Lett.} {\bf A198}, 9-13 (1995); {\it Class. Quantum Grav.} {\bf 17}, 2399-2409 (2000).

\bibitem{Carmeli}
M. Carmeli, {\it Classical Fields:  General Relativity and Gauge Theory}, World Scientific, Singapore (2001).

\bibitem{Singh3}
D. Singh, N. Mobed, and G. Papini, {\it J. Phys. A.:  Math. Gen.} {\bf 37}, 8329-8347 (2004).

\bibitem{Itzykson}
C. Itzykson and J-B. Zuber, {\it Quantum Field Theory}, McGraw-Hill, Toronto (1980).

\bibitem{Greiner}
W. Greiner, {\it Relativistic Quantum Mechanics}, Springer-Verlag, New York (1997);
J. D. Bjorken and S. D. Drell, {\it Relativistic Quantum Mechanics}, McGraw-Hill,
New York (1965).

\bibitem{Singh}
D. Singh and G. Papini, {\it Nuovo Cim.} {\bf B115}, 223-237 (2000).

\bibitem{Sakurai}
J. J. Sakurai, {\it Modern Quantum Mechanics:  Revised Edition}, Addison-Wesley,
Don Mills (1994).

\bibitem{Berry}
M. V. Berry, {\it Proc. R. Soc. Lond.} {\bf A392}, 45-57 (1984);
B. R. Holstein, {\it Topics in Advanced Quantum Mechanics}, Addison-Wesley, New York (1992).

\bibitem{Mohapatra}
R. N. Mohapatra and P. B. Pal, {\it Massive Neutrinos in Physics and Astrophysics}, World Scientific, Singapore (1991).

\bibitem{Holstein}
B. R. Holstein, {\it Am. J. Phys.} {\bf 61}, 142-147 (1993);
K. Fujikawa, {\it Phys. Rev.} {\bf D21}, 2848-2858 (1980).

\bibitem{Mielke}
E. W. Mielke, A. Mac\'{i}as, and H. A. Morales-T\'{e}cotl, {\it Phys. Lett.} {\bf A215}, 14-20 (1996);
E. W. Mielke and D. Kreimer, {\it Int. Jour. Mod. Phys.} {\bf D7}, 535-548 (1998);
E. W. Mielke, {\it Int. Jour. Theor. Phys.} {\bf 40}, 171-189 (2001);
D. Kreimer and E. W. Mielke, {\it Phys. Rev.} {\bf D63}, 048501 (2001).

\bibitem{Ramond}
P. Ramond, {\it Field Theory:  A Modern Primer, Second Edition}, Addison-Wesley, Don Mills (1990).

\bibitem{Singh2}
D. Singh, in preparation.

\bibitem{Dandoloff}
R. Dandoloff, {\it Phys. Lett.} {\bf A139}, 19-20 (1989);
R. Dandoloff and W. J. Zakrzewski, {\it J. Phys. A:  Math. Gen.} {\bf 22}, L461-L466 (1989).

\bibitem{Aharonov}
Y. Aharonov and D. Bohm, {\it Phys. Rev.} {\bf 115}, 485-491 (1959).

\bibitem{Morales}
H. A. Morales-T\'{e}cotl, A. Mac\'{i}as, and E. W. Mielke, ``Geometric Phases and Translations,'' in
{\it Recent Developments in Gravitation and Mathematical Physics:  Proceedings of the First Mexican
School on Gravitation and Mathematical Physics, Guanajunto, Mexico, 12-16 December, 1994},
Eds. A. Mac\'{i}as, T. Matos, O. Obreg\'{o}n, H. Quevedo, 242-247, World Scientific, Singapore (1995);
E. W. Mielke, J.D. McCrea, Y. Ne'eman, and F. W. Hehl, {\it Phys. Rev.} {\bf D48}, 673-679 (1993).

\end{thebibliography}
\end{document}